\documentclass{aastex}          %% The manuscript based on AASTeX v5.x
\usepackage{spr-astr-addons}    %% mimicing ASTR journal style
\begin{document}
%% Article title
%
\title{Detecting metal-rich intermediate-age Globular Clusters in NGC 4570
  using K-band photometry}

%Using Globular Clusters to derive the formation history of the Virgo Lencticular NGC4570}

%% Running heads
\shorttitle{Intermediate-age and metal-rich GCs in NGC4570}
\shortauthors{Kotulla, Fritze and Anders}

%% Author and Affilations
\author{R. Kotulla\altaffilmark{1}} 
\and 
\author{U. Fritze\altaffilmark{1}}
\and 
\author{P. Anders\altaffilmark{2}}
%\affil{}
%\email{} %% non-output

%% Alternate Affilations
\altaffiltext{1}{Centre for Astrophysics Research, University of
  Hertfordshire, College Lane, Hatfield AL10 9AB, UK}
\altaffiltext{2}{Sterrekundig Intituut, Universiteit Utrecht, Princetonplein 5, 3584 CC Utrecht, NL}
%\altaffiltext{3}{}

%% Abstract
\begin{abstract}
Globular Cluster Systems (GCSs) of most early-type galaxies feature two
peaks in their optical colour distributions. Blue-peak GCs are believed to be old and
metal-poor, whereas the ages, metallicities, and the origin of the red-peak
GCs are still being debated. We obtained deep K-band photometry and combined
it with HST observations in g and z to yield a full SED from optical to
near-infrared. This now allows us to break the age-metallicity degeneracy.

We used our evolutionary synthesis models GALEV for star clusters to compute a
large grid of models with different metallicities and a wide range of
ages. Comparing these models to our observations revealed a large population
of intermediate-age (1--3 Gyr) and metal-rich ($\approx$ solar metallicity)
globular clusters, that will give us further insights into the formation
history of this galaxy.
\end{abstract}

%% Keywords
%\keywords{}

%%  Please use labels (\label, \ref) for section, figure, table, 
%%  equation  reference. Use \cite for bibliography references.
%
\section{Introduction}\label{s:intro}
Globular Cluster Systems (GCSs) are now recognized as powerful tracers of
their parent galaxy's formation history \citep{west04,fritze04,brodie06}.
From their age and metallicity distributions one can reconstruct the parent
galaxy's (violent) star formation and chemical enrichment history all the way
from the very onset of star formation in the early universe to the present.

Most early-type galaxies show a bimodal color distribution for their GCSs
\citep[e.g.][]{gebhardt99,kundu01a,kundu01b,peng06}: A universal blue peak and
a red peak with colors and height relative to the blue peak varying from
galaxy to galaxy. The blue peak GCs are believed to be old and metal-poor, the
origin of the red peak is still unclear.

\subsection{Our approach to lift the age-metallicity degeneracy}

Optical data alone do \textbf{not} allow to disentangle ages and
metallicities: Colour-to-metallicity transformations have to assume an age,
while colour-to-age transformations are only valid for one metallicity.
This degeneracy, however, can be broken by including near-infrared data that
are more sensitive to changes of metallicity rather than age. 

\cite{anders04b} used extensive artificial star cluster tests and showed that
observations in 
\begin{itemize}
  \item 3 passbands for GCSs in dust-free E/S0s or
  \item 4 passbands for (young) star clusters in dusty environments,
  \item spanning as wide as possible a wavelength-basis (U through K) and
  \item including at least one NIR-band (e.g. K) with
  \item accuracies $\leq 0.05$ mag in the optical and $\leq 0.1$ mag in the
    NIR
\end{itemize}
allow to disentangle ages and metallicities and determine
\textbf{individual GC metallicities to $< \pm 0.2$ dex, and ages to $< \pm
  0.3$ dex}, i.e. they allow to distinguish $\leq 7\,\rm Gyr$ old GCs from those
$\geq 13\,\rm Gyr$ old.

Similar studies also using NIR-data to determine ages and metallicities of
globular clusters have only been done for a few galaxies until now
\citep{puzia02,kisslerpatig02,hempel03,larsen05,hempel07}. More than half of
these are found to host a population of GCs that is younger and/or more
metal-rich than the old and metal-poor GC population in the Milky Way.

We note that many of these previous studies relied on cumulative distribution.
The most important change compared to our analysis is that we derive the
physical parameters for each individual cluster. That also enables us to study
correlations of these parameters, e.g. with their spatial distributions.

%\subsection{Why NGC4570}

\section{Models}
We used our GALEV evolutionary synthesis models for star clusters
\citep{schulz02,anders03} to compute a large grid of models for five different
metallicities $\rm -1.7 \leq [Fe/H] \leq +0.4$ and ages between $\rm 4\,Myr$
and $\rm16\, Gyr$ with time-steps of $\rm 4\, Myr$. Since early-type galaxies
do not contain significant amounts of dust we did not include extinctions $\rm
E(B-V)>0$ into our grid. We therefore only need three filters (HST
F475W,F850LP and SOFI Ks) to determine all relevant parameters
(age, metallicity, mass) for each cluster.

\begin{figure}[tb]
\includegraphics[width=\columnwidth]{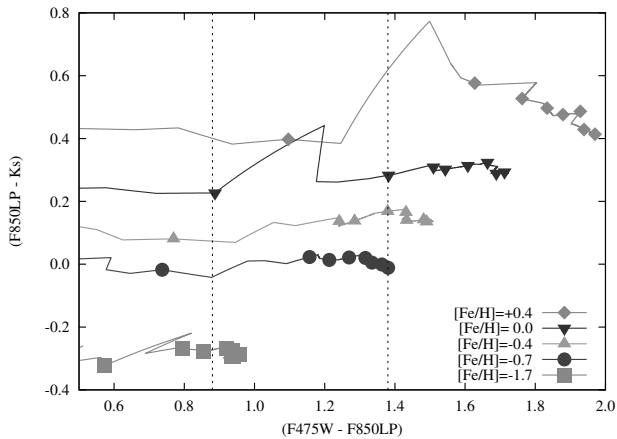}
\caption{Color-color diagram showing evolutionary tracks of
  globular clusters of different metallicity in the ${\rm
    (F850LP-K_s)-(F475W-F850LP)}$ plane, computed from our GALEV evolutionary
  synthesis models and using AB
  magnitudes; Tick-marks are shown every 2 Gyrs. The vertical lines indicate
  the peak-colours of the bimodal optical colour distribution \citep[taken
  from][]{peng06}.}
\label{fig:colorcolor}
\end{figure}

Note that we do not depend on color-transformation from the HST to Standard
Johnson filters. Our models first compute spectra as function of time, that
later are convolved with the corresponding filter curves to yield final
magnitudes. Figure \ref{fig:colorcolor} shows the resulting color-color
diagram for the three filters discussed here.

As expected, there is a degeneracy between ages and metallicities, but note how
nicely the addition of the near-infrared splits up different metallicities and
allows to break this degeneracy.

\section{Data}
\subsection{Near-infrared observations}
We used the SOFI (Son of ISAAC) near-infrared imager on the ESO-NTT during two
nights in May 2007 (ESO-program id 079.B-0511). SOFI contains a Hawaii HgCdTe
1024x1024 chip with a resolution of $\rm 0.288$ arcsec per pixel, resulting in
a field-of-view (FoV) of $\rm 4.92\times4.92\, arcmin$. This large FoV allowed
us to significantly reduce the overhead for sky-exposures using one half of
the detector for the galaxy, the other half for sky, and changing positions
every minute.

\subsection{K-band data reduction}
The data reduction was done using ESO-MIDAS and largely followed the recipes
given in the SOFI User's manual \citep{sofiusermanual}, starting with the
inter-quadrant row-crosstalk. To flat-field the data we used a combination of
dome-flats, illumination correction surfaces and a refined master-flat
obtained from normalized sky-frames of both nights. All frames were
sky-subtracted using the average of six frames taken nearest in time that have
been scaled to the sky-value of the object exposure. We aligned individual
images by matching the positions of several stars and compact background
galaxies and averaged them to give the final image.

\subsection{HST data}
We used archival data obtained from the \textit{Hubble Space Telescope}. Both
datasets (J8FS18011 and J8FS18021) were observed as part of \textit{The ACS
  Virgo Cluster Survey} \citep{cote04,jordan04} and automatically reduced and
calibrated by the On-the-fly Reprocessing (OTFR) pipeline at STScI. After
retrieval we checked the alignment of both frames relative to each other,
again using a set of stars within the ACS FoV.

\subsection{Cluster selection and photometry}
Cluster selection was done using SExtractor \citep{bertin96} requiring at
least 4 pixels with intensities above the $3\sigma$-over-background threshold.
A cross-correlation of the catalogs of both HST filters to remove remaining
spurious detections resulted in a ``optical'' catalog of $~330$ sources. For
all these cluster candidates we derived intrinsic radii using the ISHAPE
package within BALOAB \citep{larsen99}, assuming a circular-symmetric King
profile with concentration $c=30$ for the GCs.

To remove background galaxies and stars, we required all valid cluster
candidates to have radii in the range $0.2{\rm px} \leq r \leq 5 {\rm px}$
(equivalent to physical sizes of $\rm 0.8\ldots 20\,pc$ at the assumed
distance of $\rm 17\,Mpc$ \citep{tonry01} ), leaving $\approx 280$ candidates.
For all of them we performed aperture photometry with aperture sizes of 10
pixels for the HST filters and 7 pixels ($\approx 2\,\rm arcsec$) in Ks. This
allowed us to obtain (g, z, Ks) photometry for $\approx 150$ candidates; most
of the remaining cluster candidates for which we could not derive magnitudes
in all three filters were not included within the SOFI FoV.

\section{Results}
\begin{figure}[tb]
\includegraphics[width=\columnwidth]{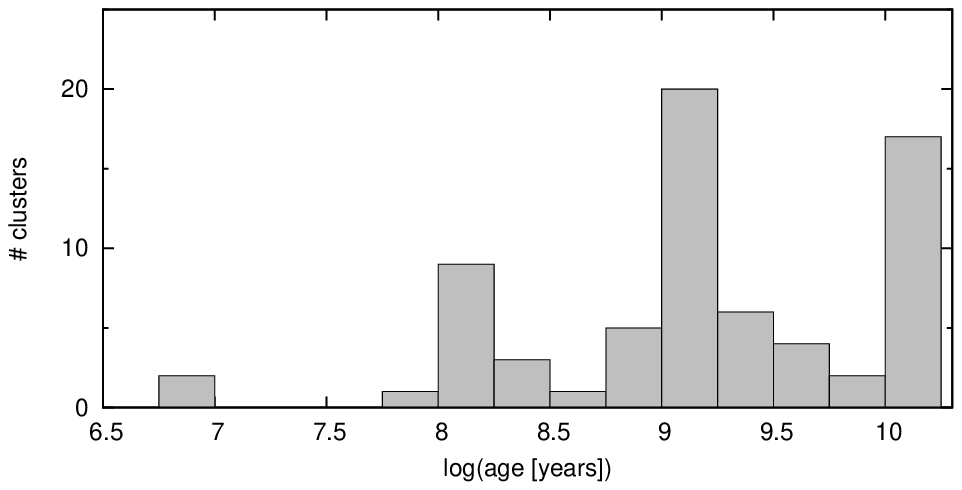}
\includegraphics[width=\columnwidth]{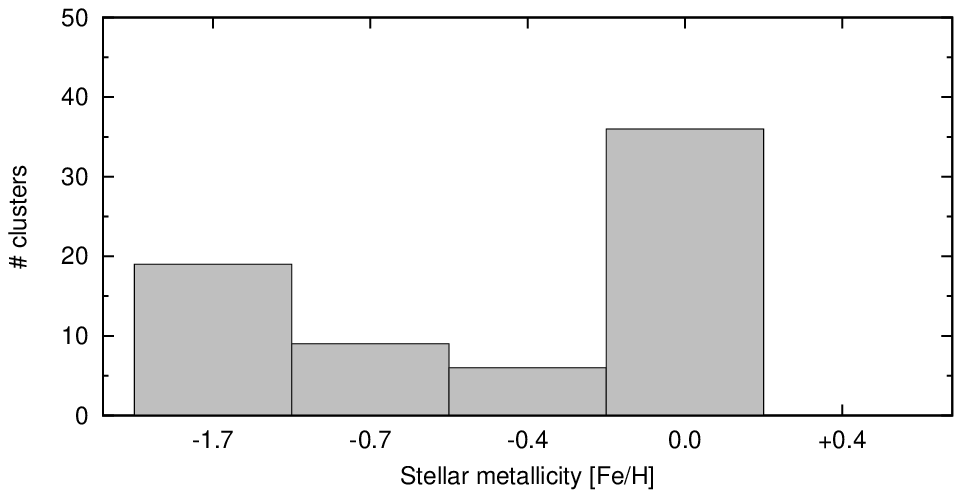}
\caption{Distribution of ages (upper panel) and metallicities (lower panel)
  for all clusters}
\label{fig:hists}
\end{figure}

We derived physical parameters for all GC candidates using AnalySED
\citep{anders04a,anders04b}. It compares the observed spectral energy
distributions (SEDs) with all model SEDs and automatically derives
probabilities for each SED on the basis of a $\chi^2$-algorithm. From these
probabilities it finds the best-fitting template and its physical parameters
age and metallicity. The results are shown in Figure \ref{fig:hists}.

\subsection{Old and intermediate ages}
The age distribution is dominated by two distinct populations: An old
population with ages older than 10 Gyr that has formed during galaxy formation
in the early universe. This population is universal in all galaxies and can be
studied in great detail within our Milky Way.  The second population has
intermediate ages of 1--3 Gyr and therefore must have formed later during a
violent episode of star-formation; the most plausible explanation for such an
event being an intense starburst as e.g. accompanying the merger of two
gas-rich spirals or the accretion of a gas-rich companion, resulting in a
phase of massive star cluster formation. Besides these two dominating
populations there is a number of clusters that do not belong to either of
these two. These can be explained by gas left over from the original galaxies
that was ejected into tidal tails and later rained down onto the merger
remnant.

A merger or accretion event is supported by the detection of a nuclear stellar
disk in the host galaxy \citep{vandenbosch98a,vandenbosch98b,scorza98}.
\cite{vandenbosch98b} estimated an age of $\leq \rm 2\, Gyr$ for the central
structure, in excellent agreement with our ages for the intermediate
population, suggesting that both have formed from the same event.

\subsection{Solar metallicities}
The lower panel of Fig. \ref{fig:hists} shows the metallicity distribution of
our cluster sample. It features an eye-catching prominent peak at solar
metallicity. 26 out of 36 of these high-metallicity clusters belong to the
intermediate-age population, however, 10 out of the 36 solar
metallicity GCs have old ages. Their origin is not yet clear.

We caution the reader that the number of old and metal-poor globular clusters
is underestimated with our analysis. Since we only include globular clusters
with a secure Ks-band detection, we are biased toward higher-metallicity
clusters, because those have significantly redder optical--near-infrared
colours.

A more detailed analysis of the properties of our cluster sample, including
their masses will be published in Kotulla et al., in prep.
%\citet[in prep.]{kotulla08}.

% In the literature (XXX) there are several scenarios that can lead to such a
% second population:

\section{Summary}
We obtained deep K-band photometry of the Virgo lenticular NGC4570 and
combined it with archival optical data from HST. We selected globular cluster
candidates based on the HST data and their intrinsic optical sizes. 

GALEV evolutionary synthesis models were used in combination with AnalySED to
automatically derive physical parameters age, metallicity and mass for each
individual cluster.

We detect a significant population of intermediate-age (1--3 Gyr) and
metal-rich ($\rm [Fe/H] > -0.4$) cluster population that has not been reported
for this galaxy before.

\acknowledgments We thank the \textit{International Space Science Institute
  (ISSI)} for their hospitality and support of this research. This publication
is based on observations made with ESO Telescopes at the La Silla Observatory
under programme ID 079.B-0511. This paper is also based on archival
observations with the NASA/ESA Hubble Space Telescope, obtained at the Space
Telescope Science Institute, which is operated by the Association of
Universities for Research in Astronomy (AURA), Inc, under NASA contract NAS
5-26555.

%%%  Using BibTeX  (Name-Year style)
%
%\bibliographystyle{spr-mp-nameyear-cnd}  %% BibTeX style
%\bibliographystyle{natbib}  %% BibTeX style
\bibliographystyle{spr-mp-nameyear-cnd} 
\bibliography{rkotulla}                %% BibTeX data

\end{document}